\documentclass[pre,reprint]{revtex4-1}
\usepackage{ae}
\usepackage[pdftex]{graphicx}
\usepackage{amsmath}
\usepackage{amssymb}
\usepackage{epsfig, float}
\usepackage{hyperref}
\usepackage{color}
\usepackage{nameref}
\usepackage{changes}

\DeclareMathOperator*{\argmin}{arg\,min}

\newcommand{\vect}[1]{\boldsymbol{#1}}

\begin{document}
\date{\today }

\title{Online human aggregation under pressure moves beyond preferential attachment}

\author
{Zhenfeng Cao$^{1}$, Minzhang Zheng$^{1}$, Pedro D. Manrique$^{1}$, Zhou He$^{2,3}$ and Neil F. Johnson$^{1}$}
\affiliation{$^1$Physics Department, University of Miami, Coral Gables, FL 33146, U.S.A.\\
$^2$School of Economics and Management, University of Chinese Academy of Sciences, Beijing 100190, China\\
$^3$Department of Industrial \& Systems Engineering, National University of Singapore, Singapore 117576}

\date{\today}

\begin{abstract}
There is a significant amount of online human activity which is either clandestine or illicit in nature, and hence where individuals operate under fear of exposure or capture. Yet there is little theoretical understanding of what models best describe the resulting dynamics. 
Here we address this gap, by analyzing the evolutionary dynamics of the supporters behind the 95 pro-ISIS online communities (i.e. self-organized social media groups) that appeared recently on a global social media site. We show that although they do not follow a conventional (i.e. size-based) preferential attachment (PA) model, their dynamical evolution can be explained by a new variant that we introduce here, which we refer to as active attraction model (AA). This AA model takes into account the locality and group heterogeneity which undoubtedly feature in humans' online behavior under pressure, but which are not contained in conventional PA models. The AA model captures both group-specific and macroscopic observations over all size ranges -- as opposed to just the tail for large groups or groups' initial growth -- suggesting that heterogeneity and locality play a crucial role in the dynamics of online extremist support. We derive approximate expressions for the group size distributions in two simple systems that involve simultaneously the mechanisms of group joining (governed by either PA or AA), group leaving, and account banning, and show how these processes influence the group size distributions. We believe this work will serve in helping understand a broad spectrum of online human activities which are either clandestine or illicit in nature, and hence where individuals operate under fear of exposure or capture. 
\end{abstract}

\maketitle


\section{Introduction}
Just as the Internet can be used for good, it also serves as an ideal vehicle for the more clandestine and illicit side of human activity. For example, extremist entities such as ISIS (known as Islamic State) make ample use of the Internet for spreading their message and propaganda materials, recruiting young people, and soliciting funds. One particular social media platform VKontakte (www.vk.com), which has $\sim 350$ million global users and is almost identical to Facebook, became the primary online social media source for ISIS propaganda and recruiting  \cite{johnson2016new}. Unlike on Facebook where pro-ISIS activity is almost immediately eliminated, support on VKontakte develops around online groups (i.e. self-organized communities) which are akin to Facebook groups that support particular everyday topics such as a sports team. These online pro-ISIS groups may not only organize premeditated attacks, but also incite decentralized lone-wolf attacks \cite{johnson2016new}. Hence, there is a compelling need to investigate the dynamics of such online groups, especially the early-growing ones whose group sizes are relatively small.\par

Here we analyze the evolutionary dynamics of the supporters behind the 95 pro-ISIS online communities (i.e. self-organized social media groups) that appeared recently on a global social media site. Though focussed on ISIS support, our model, analysis and findings should serve in helping understand a broad spectrum of online human activities which are either clandestine or illicit in nature, and hence where individuals operate under fear of exposure or capture. Indeed, outside of extremism and online illicit or clandestine activity there is already increasing interest in understanding how communities, users, or groups attract new followers/members, and develop over time \cite{backstrom2006group, mislove2007measurement, palla2007quantifying, leskovec2009community, leskovec2008statistical, traud2012social, backstrom2008preferential, berger2016occasional, kairam2012life}. Many previous studies highlight the role played by a ``non-local'' preferential attachment (PA) effect in the formation of groups or network clusters \cite{albert2002statistical, barabasi2002evolution, backstrom2006group, mislove2007measurement, leskovec2009community, kumar2010structure}. Unfortunately, we find that conventional PA models cannot well explain the unusual rapid and heterogeneous growths at the early growing stage of the online pro-ISIS groups observed from empirical data (see Sec. \ref{sec:data}), possibly due to one or more of the following reasons:
\begin{enumerate}
	\item The online pro-ISIS supporters and the self-organized groups that they form online, are under pressure. The members in such a group are discussing an extreme topic and supporting an illegal terrorist organization. These individuals have to co-habit the same online space as opposing individuals and entities (e.g. the online organization called Anonymous) as well as government agencies, all of whom are not only trying to defuse the narrative of the extremist social media groups  but are also possibly trying to track down the identities of particular group members. As a result, the extremist supporters are under continual pressure and likely act online in ways that help them maintain a more hidden profile. This challenges significantly the typical PA model assumption of implicit knowledge of all group sizes across the whole population.
	\item Extremism discussed in online pro-ISIS groups is a specialist niche topic. It seems less likely that people would be drawn to it simply because it is popular among others. Therefore, the probability of attaching should not only depend on node connectivity, but also should incorporate individual heterogeneity and locality -- just as readers of an article in a physics journal are likely to be physicists, and within that subpopulation, readers of an article on network science are more likely to be from that network science community. These are often overlooked factors that can play an important role in group formation and dynamics \cite{leskovec2008microscopic}.	
	\item The evolution of online groups is affected by moderators who have the right to ban a group. The conventional PA models fail to consider the consequence of such moderators. Their special role and powers suggest that their presence and activity can influence heavily the group evolution \cite{kelley2011predation}. While a full theory that includes them would require multi-species analysis that we cannot yet provide, it suffices to say that their presence in the system is another reason that can move the dynamics beyond PA.	
\end{enumerate}
These issues need to be addressed in order to get a more complete picture of human activity online. Here we go partway toward trying to address these issues. In so doing, our work also goes partway toward addressing the following wish-list in the network literature: (1) Many previous studies are based on particular definitions and techniques for detecting the groups. It would be very useful to have a more general framework for discussing what a group is \cite{leskovec2009community, leskovec2008statistical, yang2015defining, palla2007quantifying}. (2) Although many previous studies focus either on the observational aspects \cite{backstrom2006group, mislove2007measurement, backstrom2008preferential, berger2016occasional, kairam2012life} or theoretical modeling \cite{sorensen1987cluster, gueron1995dynamics, leskovec2008statistical}, knowledge about how network theories agree with the observation on both the microscopic and the macroscopic scales is rare. For instance, a large portion of studies focus on the global statistical properties such as the scaling behavior of the group/cluster size distribution and the evolution of the globally averaged quantities \cite{albert2002statistical, backstrom2006group, mislove2007measurement, leskovec2009community}, yet little is known about the evolution at the group-specific level. This includes the study in Ref. \cite{johnson2016new} which focuses on the mechanisms producing the tail in the size distribution at larger groups sizes. More detailed group-specific studies for any group size are needed since global statistics could be misleading, e.g.  due to the temporal variation of the global population\cite{barabasi2002evolution}.(3) Although there are studies on how microscopic node behavior would reproduce the observed macroscopic statistical properties \cite{leskovec2008microscopic}, knowledge of how individual behaviors contribute to the evolutionary property of a single group is missing.

Specifically, this paper proposes a simple growth model, namely the active attraction (AA) model, that goes beyond PA and takes into account the locality and heterogeneity of online activity. We show that this active attraction model (AA) captures both group-specific and macroscopic observations over all size ranges -- as opposed to just the tail for large groups \cite{johnson2016new} or groups' initial growth and development. Our findings suggest that heterogeneity and locality play a crucial role in the dynamics of online extremist support. We also derive approximate expressions for the group size distributions in two simple systems that involve simultaneously the mechanisms of group joining (governed by either PA or AA), group leaving, and account banning, and show how these processes influence the group size distributions. 
\par
The outline of the paper is as follows. In Sec. \ref{sec:data} we introduce the dataset and the findings. In Sec. \ref{sec:AA} we model the system by a PA mechanism, and check if the group-specific and macroscopic observations can be reproduced by the model. In Sec. \ref{sec:AA} we introduce the AA model, and show how the group-specific and macroscopic observations are reproduced by the model. In Sec. \ref{sec:Size}, we derive the analytic expressions for the group size distributions of two simple systems involving simultaneously the group joining (governed by either AA or PA), leaving, and banning processes. The conclusion is given in Sec. \ref{sec:conclusion}.

\section{Empirical data and analysis} \label{sec:data}
Our dataset is assembled using the same methodology as Ref. \cite{johnson2016new}. In contrast to Ref. \cite{johnson2016new} that focuses on the late stage when groups have become very large, we focus here on the early growing stage when the group sizes are relatively small. The dataset used in this work comprises the 95 groups identified as being pro-ISIS \cite{johnson2016new} whose dates of first appearance are within our observational period of 320 days. These provide us with detailed information about the evolution of the group memberships with a high temporal resolution up to one day. Specifically, there are three main processes involved in the group evolution: the group joining and leaving events that may occur every day during a group's lifetime, and the banning of a group by the moderators. For each group, the dataset provides information about the size, the number of joining and leaving events on each day, as well as the first appearance date which we take as being the first day on which the group has at least have one member, and the banning date if the group gets banned within the period of observation. This banning can be identified by an abrupt drop in the group size to zero.

\begin{figure}
\begin{center}
\includegraphics[width=\linewidth]{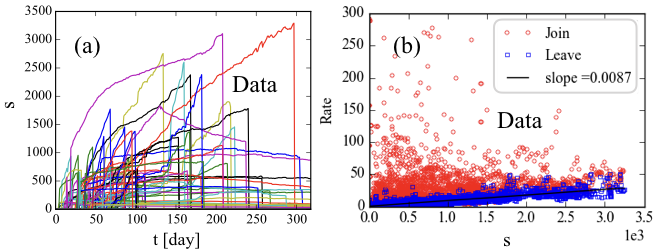}
\end{center}
\caption{(a) Evolutionary profiles of the size ($s$) of all the extremist support groups in our empirical dataset. The abrupt drop of a group's size to zero indicates the group being banned by the moderators at that instant. (b) Group joining and leaving rates vs. the group sizes from the empirical data. For the leaving rates vs. the group sizes, we also show the linear regression result.}
\label{fig1}
\end{figure}

We started by examining the evolution of the group sizes. According to PA, the size of an early-growing group should be small. However, we found that the group sizes can become very large within a few tens of days, causing the shark-fin shaped growth. This shark-fin shape is defined by the distinctive concave shape during growth (Fig. \ref{fig1}(a)).
We then studied the correlation between the group joining and leaving rates on day $t$ and the group size on that day (Fig. \ref{fig1}(b)). For a given group, the joining/leaving rate on day $t$ is estimated by applying a linear fitting to the cumulative number of joining/leaving events during the 5 days around day $t$. We carried this out for all groups and all days  -- except for the first and last 2 days, since we need 5 days around day $t$ to do the fitting. We checked that our findings are insensitive to the precise value of this time window.  We found that the group-joining data points are highly dispersed when the group size is small (we consider a group is small if its size is less than $\sim 10^3$ following Ref. \cite{johnson2016new}). This suggests that non-PA rules may apply during the early growing stage. Group-leaving data points can be well fit by a line whose slope is found to be $\sim 0.0087$ from linear regression (Fig. \ref{fig1}(b)), indicating that leaving a group is more like an independent personal decision than the act of joining a group.\par
However, this is still insufficient to exclude PA as the governing mechanism for the group joining process during the early growth stage, for the following two reasons: First, a rapid growth of the (global) total number of group members or followers may still result in decelerating growth during the early growth stage \cite{barabasi2002evolution}. Second, it could be that the temporal fluctuations in the total number of followers caused the dispersion of data points observed in Fig. \ref{fig1}(b). Therefore in the next section we attempt to test out how well PA reproduces the observations, by assuming that PA governs the group-joining process.

\section{Preferential Attachment (PA) model} \label{sec:PA}
\subsection{Size evolution of a single group}
\begin{figure}[h]
\begin{center}
\includegraphics[width=\linewidth]{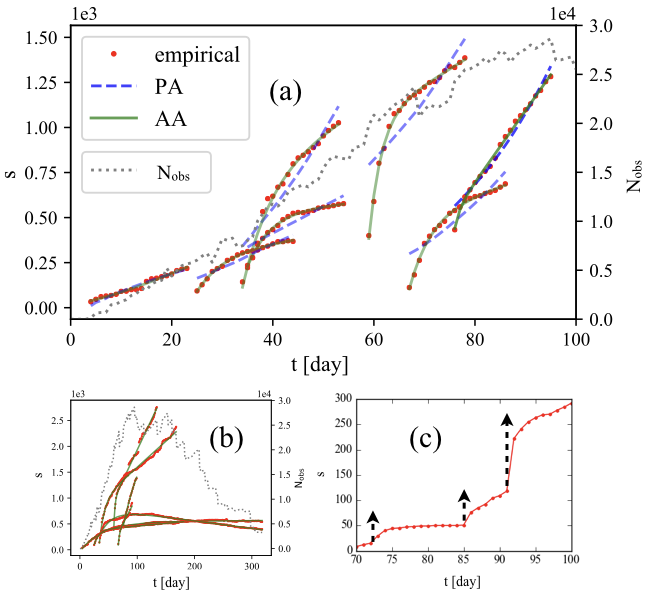}
\end{center}
\caption{(a) Model fittings for the size evolution of 7 representative extremist online groups. PA and AA corresponds to fitting by Eq. \ref{eqn:pa} and \ref{eqn:aa}, respectively. Legends in (a) are applicable to all other sub-figures. For each group in (a) only the first 20 days on which the group size is nonzero are fit. (b) shows how well the groups in (a) are fit by the AA model for their entire lifetime.  In (a)-(b), $N_{obs}$ is the evolution of the empirical global total number of followers. (c) An example of the stair-like pattern of a group's size evolution profile. The instants when the group is exposed to a new population of potential members are marked by the dashed black arrows.}
\label{fig2}
\end{figure}

We first model the size evolution of a single group. According to PA, the increase of the group size on day $t$ (before the group gets banned) is given by 
\begin{equation}\label{eqn:pa}
s[t]-s[t-1]=\frac{\alpha s[t-1]}{N_{obs}[t-1]} -\eta s[t-1],
\end{equation}
where $\alpha$ and $\eta$ describe the group joining and leaving rates, respectively, and $N_{obs}[t]$ is the total number of followers/members in the system on day $t$ (shown in Fig. \ref{fig2}(a)-(b)). We define $t_0$ as the group's appearance date, i.e., the first day that the group's size is nonzero. Given $\alpha$ and initial size $s[t_0]$, $\eta$ is estimated to be approximately $0.0087$ (Fig. \ref{fig1}(b)). Since $N_{obs}[t]$ can be directly obtained from the data, we can iteratively estimate its size on all future days. Hence the curve fitting problem is to find the optimal $\alpha$ and $s[t_0]$ that minimize the Pearson's $\chi^2$:\begin{equation}
\argmin_{\alpha ,\ s[t_0]} \sum_{t=t_0}^{t_b}\bigg(\frac{s[t]-s'[t]}{s'[t]}\bigg)^{2},
\end{equation}
where $t_b$ is the day when the group gets banned, $s[t]$ is the group size on day $t$ estimated using the iterative expression (i.e. Eq. \ref{eqn:pa}), and $s'[t]$ is its corresponding observed one. The minimization can be carried out by conventional multi-variable optimization algorithms. Note that Eq. \ref{eqn:pa} is only valid during the early growth stage when the saturation effect (i.e. the effect of finite population) can be ignored. Hence to do the fitting for each group in the dataset, we only used the sizes of the first 20 days when the group size is nonzero. 

We find that the best estimates of $\alpha$ are spread across a broad range, but the fittings for most of the groups are still poor (e.g. see Fig. \ref{fig2}(a) where the fittings barely reproduce the concave shapes of the empirical profiles). This indicates that the empirical growth rate of $N_{obs}[t]$ is far from sufficient to bring about the observed shark-fin shapes. Also against the PA is the fact that, rigorously speaking, conventional PA models implicitly assume a constant $\alpha$ for all groups. 

\subsection{Stochastic simulation of PA}
We simulated the group growth using PA as follows. We set all the parameters (including the total number of groups, the creation and banning date, the total number of new joining events on each day, etc.) to be the same as the data, except that 
\begin{enumerate}
\item we redistribute the new group joining events observed on each day to all the non-banned groups following the PA rule (i.e. the probability that a user joins a group on day $t$ is proportional to the size of the group on day $t-1$, $s[t-1]$); 
\item we use the constant group leaving rate ($\sim 0.0087$) estimated from Fig. \ref{fig1}(b);
\item for a group that appears for the first time on day $t_0$, we manually assign a small initial size $s[t_0-1]$ to it, e.g. 1. We also tested other larger values, but the main results are the same.
\end{enumerate}
The exact steps in the simulation are as follows. On day $t$, 
we first detect from the dataset which groups exist (i.e. have at least one member) and denote them by a set, $\mathcal{G}$. We also obtain directly from the dataset, the total number of new joining events  ($\Delta N_{J}[t]$). Next we redistribute the new joining events to the alive groups by drawing a sample from the multinomial distribution, $Multinomial(\Delta N_{J}[t], \vect{W})$, where $\vect{W}$ is a vector of probabilities whose values sum up to 1 and are proportional to the sizes of the groups on the previous day, By so doing, we ensure that the number of new joining events for a group is always proportional to its current size, and the total number of new joining events is the same as the data. Finally for each group, we consider the group leaving events by subtracting a value from its previous size. The value is drawn from the Binomial distribution $Binomial(s[t-1], 0.0087)$, where $s[t-1]$ is the size of the group on the previous day.

\begin{figure}[h]
\begin{center}
\includegraphics[width=\linewidth]{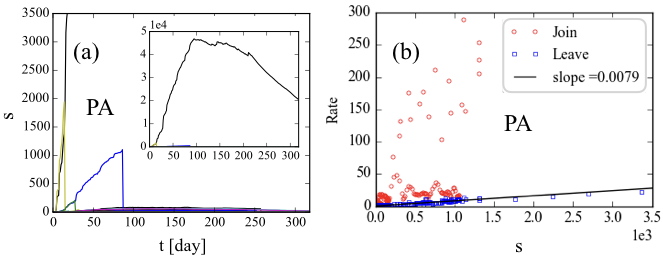}
\end{center}
\caption{(a) Evolutionary profiles of the size, $s$, of all the groups in the simulation of the PA model, where the abrupt drops of the group sizes to zero are due to the group being banned by the moderators. (b) Group joining and leaving rates vs. group sizes ($s$) of the simulation. For the leaving rates vs. the group sizes, we also show the linear regression result.}
\label{fig3}
\end{figure}

Based on the simulation results, we are also able to compare the probability density function (PDF) of the group sizes in the simulation to that in the empirical observation. The PDFs for both cases are obtained as follows. First, we record the sizes of all the groups on each day from the 80-100th day to a list $S$. We choose these days when doing the statistics because they correspond to the mature stage of the whole system, and hence have the maximum total number of groups. We also tested other days around and reduced the number of days, but the results are similar. Then we use the \emph{Fit} module in the well-known \emph{Python} package \emph{powerlaw}\cite{alstott2014powerlaw} to obtain the PDF. 

We find from this PA simulation that (1) the growths of most of the groups that are created at a later point in time, are effectively suppressed by a couple of groups that were created earlier (Fig. \ref{fig1}(c)). (2) The group-joining data points in Fig. \ref{fig1}(d) are much less dispersed than those in the empirical observation (Fig. \ref{fig1}(b)). (3) The PA-based size distribution doesn't agree well with the data. In short, neither the group-specific nor the macroscopic observations can be well reproduced by a PA model.

\begin{figure}[h]
\begin{center}
\includegraphics[width=\linewidth]{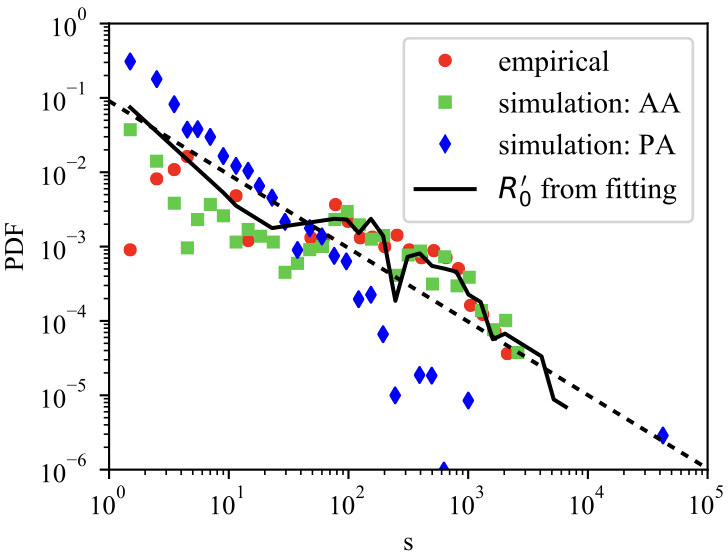}
\end{center}
\caption{Comparison between the probability density functions (PDFs) of the group sizes from the empirical observations and the simulations of the AA and PA models, and the fitting parameter $R_{0}'$ estimated from the curve fittings (from Eq. \ref{eqn:aa}), for all the groups in the empirical dataset. The black dashed line has a slope of $-1$, and serves simply as a guide to the eye.}
\label{fig4}
\end{figure}

\section{Active Attraction (AA) model} \label{sec:AA}
\subsection{Size evolution of a single group}
Here we introduce the AA model. The AA starts by making two reasonable assumptions:
\begin{enumerate}
\item Public group events (e.g. a rapid sharing of an interesting post, or an invitation letter simultaneously sent to a large population, etc.) can attract users to join quickly even when the group size is still small.
\item Groups are heterogeneous in that a given group may not be easily accessible to all users, or it may not be of equal interest to all users. 
\end{enumerate}
The first assumption can introduce some non-PA effects, while the second stresses the locality and heterogeneity in user-group interactions. With these two assumptions, one can imagine observing a stair-like growth in group size (e.g. Fig. \ref{fig2}(c)) if several major public group events occur in turn. This motivates our new AA model beyond PA, and its name which reflects a more specific user-group active attraction.
In particular, based on the fact that the groups in our dataset are under pressure and the topic itself will only appeal to an extreme fraction of the population of online users, a group may be accessible or of interest to only a portion of the whole population. This portion is determined by a saturation level denoted by $R$. For simplicity, we assume $R$ changes linearly with time, i.e. $R = R_{0} + \gamma (t-t_0)$, where $R_{0}$ is the initial size, and $\gamma$ is the rate of change. Next, we define a joining rate ($0<\alpha<1$ constant in time,  but could be different for each group), a leaving rate ($\eta$, as mentioned earlier and found to be approximately $0.0087$ for all groups), and an average (over all groups) group-banning rate $\beta$, which is only used in the simulation. With these settings, the evolution of the size of a single group before banning (hence no $\beta$ dependence) in the continuous limit is given by 
\begin{equation}
\begin{split}
\frac{ds(t)}{dt}&=\alpha  [R-s(t)]-\eta s(t)\\
&=\alpha  [R_0+\gamma  (t-t_0)-s(t)]-\eta s(t),
\end{split}
\end{equation} 
and $s(t_0)=0$. We further define $\alpha' = \alpha + \eta$ as the effective joining rate, and $R' = \alpha R/ (\alpha + \eta)$ as the effective saturation level. Let $R_0'$ be $\alpha R_0/(\alpha + \eta)$, and $\gamma'$ be $\alpha \gamma /(\alpha +\eta)$. Then the differential equation can be rewritten as $ds(t)/dt=\alpha'  [R'-s(t)]=\alpha'  [R_0'+\gamma'  (t-t_0)-s(t)]$, whose solution is given by 
\begin{equation}\label{eqn:aa}
s(t) = (R_0' -\frac{\gamma'}{\alpha'}) [1-e^{-\alpha'(t-t_0)}]+\gamma'(t-t_0),
\end{equation}
which contains 4 free parameters (i.e., $t_0$, $R_0'$, $\alpha'$ and $\gamma'$), and is the expression used for the fittings by the AA model shown in Fig. \ref{fig2}(a)-(b). The fitting is done using a conventional multi-variable optimization algorithm. Specifically, we used the \emph{curve\_fit} function with bounds set properly in the well-known \emph{Python} package \emph{SciPy}\cite{scipy}. 

We find that Eq. \ref{eqn:aa} fits most of the shark-fin group growth profiles well, not only in terms of their early growth stage (e.g. Fig. \ref{fig2}(a)), but also over their entire lifetime in many cases (e.g. Fig. \ref{fig2}(b)). In addition, we find that there exists a high heterogeneity in the fitting parameters. That is, $R_0'$, $\alpha'$ and $|\gamma'|$ ($\gamma'$ can be either positive or negative) all spread over a broad range -- and indeed, they scale roughly like a power-law with the power-law exponent $\sim 1$, except $t_0$ which is not very sensitive and is always just a couple of days earlier than the observed first-appearing day. We also tried to fix the three parameters (i.e., $R_0'$, $\alpha'$ and $\gamma'$) to a reasonable constant value (e.g., the median or mean value from previous fittings that allowed them to change), but found it resulted in poorer fittings. Hence, for the empirical data, the heterogeneity indeed exists in all these three parameters. Since the distributions of these three parameters are not the focus of this work, we show only the distribution of $R_0'$ in Fig. \ref{fig4}, which is used in the simulations. 

\subsection{Stochastic simulation of AA}
We now show how we performed the simulation of the AA model. Similar to the simulation of the PA model, we controlled all the parameters to be the same as the dataset, except that (1) we adopt a constant group-leaving rate of $\sim 0.0087$; (2) we redistribute the new joining events on each day among all the alive groups on that day according to the AA rule  (i.e. the probability that a user joins a group on day $t$ is proportional to $R[t-1]-s[t-1]$, where $R[t-1]$ and $s[t-1]$ are the saturation level and the group size on day $t-1$, respectively); (3) we further assume $\gamma=0$ for each group (and hence $R$ is constant), and set the saturation level of each group to $R_0$ ($R_0 = (\alpha + \eta) R_0')/\alpha$) estimated from the curve fitting by Eq. \ref{eqn:aa}. Note that the AA rule we used implicitly assumes $\alpha$ is a constant for all the groups -- otherwise the probability of joining a group should be proportional to $\alpha(R[t-1]-s[t-1])$ -- and hence the simulation focuses on studying the effect of the heterogeneity in the saturation level. We could have also set the $\alpha$ of each group to be the value obtained from the fitting, which would make our simulation agree even better with the dataset, but our simulations show that the heterogeneity in $R_0$ is sufficient to reproduce the empirical observations. We did the simulation in a way that is similar to the PA simulations, as follows: On day $t$, we first detect from the dataset which groups are alive (i.e. have at least one member) and denote them by a set, $\mathcal{G}$, and also obtain directly from the dataset the total number of new joining ($\Delta N_{J}[t]$) events. Next we redistribute the new joining events to the alive groups by drawing a sample from the multinomial distribution, $Multinomial(\Delta N_{J}[t], \vect{W})$, where $\vect{W}$ is a vector of probabilities whose values sum up to 1 and are proportional to $\alpha(R[t-1]-s[t-1])$. Finally for each group we consider the group leaving events by subtracting a value from its previous size; the value is drew from the Binomial distribution $Binomial(s[t-1], 0.0087)$.

\begin{figure}[h]
\begin{center}
\includegraphics[width=\linewidth]{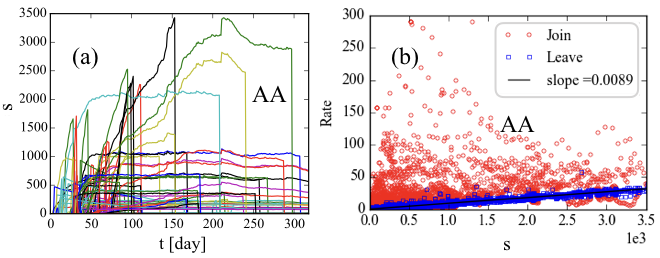}
\end{center}
\caption{(a) Evolutionary profiles of the size, $s$, of all the groups in the simulation of the AA model, where the abrupt drops of the group sizes to zero are due to the group being banned by the moderators. (b) Group joining and leaving rates vs. group sizes ($s$) for the simulation. For the leaving rates vs. the group sizes, we also show the linear regression result.}
\label{fig5}
\end{figure}

Though the heterogeneity in $\alpha$ and $\gamma$ is neglected, we find that the simulation can reproduce well the shark-fin shapes (Fig. \ref{fig5}(a)), and also the dispersion patterns in the joining and leaving rates (Fig. \ref{fig5}(b)), as well as the group size distribution (Fig. \ref{fig4}).

\section{Theoretical analysis of the group size distribution} \label{sec:Size}
We now study mathematically the group size distribution of such a system involving simultaneously the three mechanisms of group joining, leaving, and banning due to moderators. Since similar systems can exist in many other areas of human extremist activity, we cast our discussion in a more general sense and show the analytical results for both the case when the group joining follows the AA rule, and that when it follows the PA rule. Conventional rigorous treatment of the problem may involve solving the master equations, but become almost intractable -- especially when taking into account the heterogeneity of the coefficients. Hence, we resort to an approximative method and study only two simple cases. Here are the details of the derivations. 

\subsection{The AA case}
We start with the case of the group joining being an AA process. For simplicity, we consider the case that the groups are created with the same probability on each day during the periods of observation; hence the probability that a group created on day $t_0$ will remain alive (or observable) on day $t$ is $\sim (1-\beta)^{t_{\Delta}}$, where $\beta$ is the probability that a group will get banned on a day, and $t_{\Delta} := t-t_0$. As mentioned earlier, in the case that the joining is an AA process then the size evolution of a group, $s(t)$ (hereafter, we omit its $t$ label for convenience), is governed by $ds/dt = \alpha' (R' -s)$. For simplicity, we consider the case when $R'$ and $\alpha'$ are both time-independent. In such a case, the solution of this differential equation when the initial condition $s(t_0)=0$ is given by $s = R' (1- e^{-\alpha' t_{\Delta}})$, from which we can inversely get $t_{\Delta}= -\ln(1-s/R')/\alpha'$, and hence $dt_{\Delta}/ds = 1/[\alpha' (R'- s)]$. Consider the distribution of $R_0'$ in Fig. \ref{fig4}, the detailed shape of which ( though not the main concern of this work) is probably irregular due to poor statistics or the irregular banning events of the moderators etc., but which very roughly distributes around a power-law whose exponent is $\sim 1$. Inspired by this shape, we consider the simple case that the effective saturation level $R'$ of all the groups follows a power-law distribution of $P_{R'}(R') \sim R'^{-\lambda}$, and that $a'$ is the same for all groups. Then by ignoring the stochastic fluctuations in the group size evolution profiles (meaning that there is now a one-to-one correspondence between $t_{\Delta}$ and $s$ for a given $R'$), the probability density that a group created at $t_0$ is observed at $t$ having size $s$ is given by $P_{AA}(s)\sim \int_{s}^{\infty} (1-\beta)^{t_\Delta} P_{R'}(R') (dt_{\Delta}/ds) dR'  \sim \int_{s}^{\infty} (1-s/R')^{-\ln(1-\beta)/\alpha'-1} R'^{-(\lambda +1)}dR'$. Expanding the integrand with respect to $s/R'$ and keeping the terms up to $O(s/R')$, we obtain $P_{AA}(s) \sim \int_{s}^{\infty}  \{1+[ \ln(1-\beta)/\alpha'+1]s/R' \}R'^{-\lambda-1} dR' \sim s^{-\lambda}$, i.e. it also follows a power-law distribution that scales approximately the same way with the distribution of $R'$. This similarity between the saturation level distribution and the  group size distribution is also found in the empirical observation (Fig. \ref{fig4}). 

\subsection{The PA case}
For comparison, we also calculate in a similar way the group size distribution for the case that the group joining is a PA process. Neglecting the finite-population effect and considering only the case when $N(t)$ is a constant, we have $ds/dt \sim (\alpha -\eta)s$, whose solution $s\sim e^{t_{\Delta}/(\alpha-\eta)}$. Hence $t_{\Delta} \sim \ln(s)/(\alpha -\eta)$ and $dt_{\Delta}/ds \sim 1/[s(\alpha-\eta)]$. Therefore, the probability density of the group size is given by $ P_{PA}(s) \sim (1-\beta)^{t_{\Delta}} dt_{\Delta}/ds \sim s^{-\nu}$, where $\nu = -\ln(1-\beta)/(\alpha -\eta) +1$, which is greater than 1 for any $0<\beta<1$ and $\alpha > \eta$.

\subsection{Numerical verification and discussion}
We now check these analytical results since their derivations involve several approximations. We conducted stochastic simulations for both systems. The simulation settings comply to the basic assumptions we made when defining the systems for the analytic derivations. 
For the AA case, we initialize 1000 groups by assigning to each of them an initial group size of 1 and a saturation level ($R$) sampled from a Zipf's distribution with a power-law exponent equaling $\lambda$. Next, on each day $t$ we generate the number of joining events for each group by sampling from the binomial distribution $Binomial(R-s[t-1], \alpha)$ and add it to $s[t-1]$ to obtain $s[t]$, where $\alpha$ is the joining rate. Then we generate the number of leaving events for each group by sampling from $Binomial(s[t-1], \eta)$, where $\eta$ is the leaving rate, and without loss of generality was set to be $0.0087$. Finally we ban each group with a probability $\beta$ (without loss of generality, we use a value of 0.02) by setting the number of members to 1, which means after banning, we immediately create another group so that the total number of groups is always a constant.

\begin{figure}[h]
\begin{center}
\includegraphics[width=\linewidth]{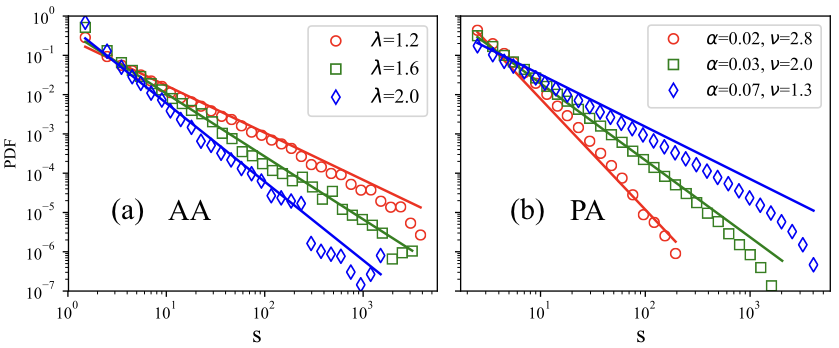}
\end{center}
\caption{Group size distributions for the (a) AA and the (b) PA model. Plot shows simulation results (dots) vs. analytic results (solid lines, which have the same color as the corresponding simulation result. Results for different parameters are shown. }
\label{fig6}
\end{figure}

We run for a sufficiently large number (e.g. 2000) of steps (i.e. days) for each run. This means that the creation dates of the groups are effectively randomized automatically after a long period due to the randomness in the group banning, and the total number of followers will also converge to a stable level. In addition, to improve the precision, we repeat the simulation 10 times for every set of parameters and then obtain the average distributions by combining the sample from each run. For each run, without loss of generality, the sample is formed by the group sizes on each day during the 80-100th day of the last 320 days in the simulation.

With respect to the simulation for the PA case, everything is the same as for the AA case, except that we don't need $R'$ anymore; on each day $t$, the first step in the AA case (see above) is to generate the number of joining events for each group; the number is sampled from the binomial distribution $Binomial(s[t-1], \alpha)$ and is added to $s[t-1]$ to obtain $s[t]$.

Figure \ref{fig6} shows that our analytical results agree well with the stochastic simulations. We can see that for the AA model, the distribution of the saturation level plays a pivotal role in determining the group size distribution. By contrast, for the PA model all three mechanisms matter. In addition for the PA case, when the group banning rate is small and the group leaving rate is significantly smaller than the joining rate, the power-law exponent is always around 1. We can also see that different microscopic user-group interaction mechanisms (e.g. AA and PA) may result in the same power-law exponent in the macroscopic statistics, which means that how a macroscopic quantity scales could be insufficient to tell the microscopic mechanism. Hence user and group level analysis -- as presented in this paper -- becomes important.

\section{Conclusion}\label{sec:conclusion}

In conclusion, we have shown that the non-PA effect is crucial for explaining the rapid growth of groups observed during the early growth stage. The PA effect then becomes more appropriate when the group size is large, as indicated in Ref. \cite{johnson2016new}). We proposed a simple non-PA model that catches the effect of locality and microscopic heterogeneity in the dynamics of group formation, and which are overlooked by conventional PA models. While we concede that there could be alternative models, our model is arguably one of the simplest, and it explains well both the group-specific and global statistical observations. Since the AA process is size-independent, it provides a novel and reasonable explanation for the cascading joining that results in a shark-fin shape observed in group size evolution. Such an observation can barely be described by a PA model without introducing an unusually high growth rate for the global total number of followers. In addition, since both PA and AA could produce a similar group size distribution, this work suggests the importance of a deeper understanding of the behaviors of individual users and groups.

There are still many open questions. For instance, the origin of the user and group level heterogeneity is still a mystery. More specifically, it is not clear if the broad distribution of the saturation levels originates from the topology of the network that influences the accessibility of the groups (i.e. the locality), or is due to some other more complicated mechanisms that result in the heterogeneity in users' interests in the groups. Hopefully the present work draws researchers' attention to these open questions, and serve as a stepping-stone toward answering them.

{\bf Acknowledgments} 
We are grateful to Yulia Vorobyeva, Andrew Gabriel and Anastasia Kuz for initial help with data collection and analysis. NFJ gratefully acknowledges funding under National Science Foundation (NSF) grant CNS1522693 and Air Force (AFOSR) grant FA9550-16-1-0247. The views and conclusions contained herein are solely those of the authors and do not represent official policies or endorsements by any of the entities named in this paper. 


\bibliographystyle{apsrev4-1} 
\bibliography{bibs} 

\end{document}